\documentclass[aps,twocolumn,prb,amsmath,superscriptaddress,nobibnotes]{revtex4}
\usepackage{graphicx}
\graphicspath{{Figures/}}
\usepackage{physics}
\usepackage{siunitx}
\usepackage{dcolumn} 
\usepackage{bm}
\usepackage{epsfig}
\usepackage{pifont}
\usepackage{natbib}
\usepackage{bibentry}
\usepackage{datatool}
\usepackage{notes2bib}
\usepackage{patchcmd}
\usepackage{physics}
\usepackage[linktoc=all]{hyperref}
\usepackage{cleveref}
\usepackage{amssymb}
\usepackage{float}

\def\laco*{LaCoO$_3$}
\def\ehs*{$\epsilon_{\text{HS}}$}
\def\eis*{$\epsilon_{\text{IS}}$}
\def\els*{$\epsilon_{\text{LS}}$}
\newcommand{\bM}{\mathbf{M}}
\newcommand{\bP}{\mathbf{\widetilde{M}}}
\newcommand{\bR}{\mathbf{R}}
\newcommand{\bI}{\mathbf{I}}

\newcommand{\bk}{\mathbf{k}}

\newcommand{\mk}{\mathbf{m}_{\mathbf{k}}}

\def\kha*{Khaliullin}
\def\s*{$|s\rangle$}
\def\t*{$|t_\alpha\rangle$}
\def\se*{$s$}
\def\jo*{$j_{1/2}$}
\def\jt*{$j_{3/2}$}
\newcommand{\bex}{\be_{x}}
\newcommand{\re}{\operatorname{Re}}
\newcommand{\im}{\operatorname{Im}}
\def\te*{$t_{\alpha}$}

\newcommand{\bi}{\mathbf{i}}

\newcommand{\bh}{\mathbf{h}}
\newcommand{\benu}{\be_{\nu}}
\newcommand{\be}{\mathbf{e}}

\newcommand{\bphi}{\bm{\phi}}

\allowdisplaybreaks

\begin{document}
\title{Excitonic magnet in external field: complex order parameter and spin currents}

\author{D. Geffroy}
\affiliation{Department of Condensed Matter Physics, Faculty of
  Science, Masaryk University, Kotl\'a\v{r}sk\'a 2, 611 37 Brno, Czechia}
\affiliation{Institute for Solid State Physics, TU Wien, 1040 Vienna, Austria}
\author{A. Hariki}
\affiliation{Institute for Solid State Physics, TU Wien, 1040 Vienna, Austria}
\author{J. Kune\v{s}}
\email{kunes@ifp.tuwien.ac.at}
\affiliation{Institute for Solid State Physics, TU Wien, 1040 Vienna, Austria}
\affiliation{Institute of Physics, CAS, Na Slovance 2, 182 21 Praha 8, Czechia}
\date{\today}

\begin{abstract}
  We investigate spin-triplet exciton condensation in the two-orbital
  Hubbard model close to half filling by means of dynamical mean-field theory. 
  Employing an impurity solver that handles complex off-diagonal hybridization
  functions, we study the behavior of excitonic condensate in 
  stoichiometric and doped systems subject to external magnetic
  field. We find a general tendency of the triplet order parameter to lay
  perpendicular with the applied field and identify exceptions from this
  rule. For solutions exhibiting $\bk$-odd spin textures, we discuss the
  Bloch theorem which, in the absence of spin-orbit coupling, forbids
  the appearance of spontaneous net spin current. We demonstrate that
  the Bloch theorem is not obeyed by the dynamical mean-field theory.

\end{abstract}
\maketitle
\section{introduction}
In 1961, N. Mott~\cite{mott61} proposed  that
the condensation of electron-hole pairs could lead to a new state 
of matter, the excitonic insulator. Subsequent theoretical
studies~\cite{knox63,keldysh65,desc65,halperin68b}
revealed a rich spectrum of possible excitonic phases. Recently,
several materials were proposed to exhibit excitonic
condensation~\cite{Jain2017}, however, unambiguous experimental proof of
excitonic condensate is available only for bi-layer quantum well
systems.~\cite{eisenstein04} In tightly bound excitons the
ferromagnetic Hund's exchange favors triplet $S=1$ over spin-singlet
state. Their condensation gives rise to several states with broken
spin isotropy ~\cite{halperin68b,balents00b,kunes14c}.

Spin-triplet exciton condensates were investigated both in
models~\cite{khaliullin13,kaneko14, kunes14c, kaneko15, Tatsuno2016,
  nasu16, Kunes2016} and material specific
studies~\cite{kunes14b,Yamaguchi2017,Afonso17}. A combination of doping
and various hopping patterns in the two-orbital Hubbard 
model was used~\cite{Kunes2016} to obtain excitonic phases
that exhibit a net magnetic polarization, antiferromagnetic spin-wave 
structures or reciprocal space spin textures.
In this Article we investigate the effect of an external magnetic field 
on these states. While the behavior of the ferromagnetic exciton
condensate (FMEC) is obvious, the response of states with no net
magnetization is less clear and is studied using the dynamical
mean-field theory (DMFT). Calculations are performed for
a spin-isotropic SU(2) model, allowing for complex off-diagonal
hybridization functions in the auxiliary impurity problem.

Particular attention is paid to the response of the state with $p$-wave
spin texture, which arises due to a dynamically generated spin-orbit
(SO) entanglement~\cite{Kunes2016}. The SO entanglement, usually
due to intrinsic SO coupling, is 
a prerequisite for the control of spin polarization by charge currents 
and vice versa~\cite{Wadley2016}. The SO entanglement generated 
by spontaneous symmetry breaking~\cite{Wu2007,Kunes2016} is little explored.
The breaking of the inversion symmetry and of the spin-isotropy in the state
with $p$-wave spin texture allows the existence of a net spin current 
in the system. However, the existence of a spontaneous net spin current
in the ground state or in thermal equilibrium is forbidden by a variational
principle~\cite{}. It is therefore interesting to find out whether 
this is obeyed by DMFT. The investigation of spontaneous currents in
the ground state of quantum systems has a long history, in the context
of superconductivity\cite{To1938, Smith1935, Bohm1949}, exciton
condensation~\cite{Volkov1978,Volkov1981,Gorbatsevich1983,
  Tugushev1984}, and systems of charged particles in the presence of
an external field\cite{Ohashi1996}. Recently, spontaneous currents in
bilayer graphene\cite{Jung2015} and superconducting systems with SO
coupling\cite{Mironov2017} were studied.

The paper is organized as follows. In Sec.~\ref{sec:model} we
introduce the computational technique and the studied observables. In
Sec.~\ref{sec:h0} we study in detail the evolution of the order
parameter across the different excitonic phases. In Sec~\ref{sec:h} we
investigate the behavior of the excitonic condensate in a magnetic
field. In Sec~\ref{sec:pheno} we interpret the numerical results using
a Ginzburg-Landau functional. Finally, we investigate the presence of
spin current in the state with $p$-wave spin texture.

\section{Model and method}
\label{sec:model}
\subsection{CT-QMC with complex hybridization}
We consider the two-band Hubbard model with nearest-neighbor (NN)
hopping on a bipartite (square) lattice. The model Hamiltonian is given by
\begin{equation}
  \label{eq:model}
  H = H_{\text{t}} + H_{\text{loc}} + H_{\text{ext}},
\end{equation}
with
\begin{equation}
  \begin{aligned}
    \label{eq:2bhm}
    H_{\text{t}} &= \sum_{\nu=x,y}\left(T_{\nu}^{\phantom\dagger}+T_{\nu}^{\dagger}\right),\\
    T_{\nu}&=\sum_{\bi,\sigma} \left(
      t_a a_{\bi+\benu\sigma}^{\dagger}a^{\phantom\dagger}_{\bi\sigma}+
      t_b b_{\bi+\benu\sigma}^{\dagger}b^{\phantom\dagger}_{\bi\sigma} \right.\\
    &\left. \quad \quad \quad +V_{+\nu}a_{\bi+\benu\sigma}^{\dagger}b^{\phantom\dagger}_{\bi\sigma}+
      V_{-\nu}b_{\bi+\benu\sigma}^{\dagger}a^{\phantom\dagger}_{\bi\sigma}
    \right),\\
    H_{\text{loc}}&=\frac{\Delta_{\text{CF}}}{2}\sum_{\bi,\sigma} \left(n^a_{i\sigma}-n^b_{i\sigma}\right)\\
    &+ U \sum_{\bi}\left(n^a_{\bi\uparrow}n^a_{\bi\downarrow}+n^b_{\bi\uparrow}n^b_{\bi\downarrow}\right)+
    U'\sum_{\bi,\sigma\sigma'} n^a_{\bi\sigma}n^b_{\bi\sigma'}\\
    &-J\sum_{\bi,\sigma} \left(n^a_{\bi\sigma}n^b_{\bi\sigma} 
      +\lambda
      a_{\bi\sigma}^{\dagger}a_{\bi-\sigma}^{\phantom\dagger}b_{\bi-\sigma}^{\dagger}b_{\bi\sigma}^{\phantom\dagger}
    \right) \\
    H_{\text{ext}}&=-\sum\limits_{\bi,\alpha \beta} 
    \bm{h}\cdot\bm{\sigma}_{\alpha \beta} 
    \left(a^\dagger_{\bi \alpha} a^{\phantom \dagger}_{\bi \beta} + 
      b^\dagger_{\bi \alpha} b^{\phantom \dagger}_{\bi \beta} \right).\\
  \end{aligned}
\end{equation}
Here, $\benu$ stands for the lattice vector of the 2D square lattice,
$c^{\dag}_{\bi\sigma}$ ($c_{\bi\sigma}$) ($c=a,b$) are the creation
(annihilation) operators with spin $\sigma$ at site $\bi$ and
$n^{c}_{\bi,\sigma} \equiv c^{\dag}_{\bi\sigma}c_{\bi\sigma}$.
The kinetic part $H_{\rm t}$ includes NN hopping between identical
orbitals $t_a$ and $t_b$, as well as cross-hopping between 
different orbitals $V_{\pm\nu}$. We are going
to study cross-hopping $(V_x,V_y,V_{-x},V_{-y})$ with 
fixed amplitude $V$ and various sign patterns:
$s$-wave $++++$, $p$-wave $++--$ and $d$-wave $+-+-$. 
The local part $H_{\text{loc}}$ contains the crystal-field splitting
$\Delta_{\text{CF}}$ between orbitals $a$ and $b$, the Hubbard
interaction $U$ and Hund's exchange $J$. The parameters
$\Delta_{\text{CF}}$ and $J$ are chosen such that the system is in the
vicinity of the low-spin (LS) and high-spin (HS) transition~\cite{werner07,Kunes2014b}.
$H_{\text{ext}}$ describes the coupling to the external
magnetic (Zeeman) field $\bf{h}$. We will present the results of
calculations performed in the density-density approximation
($\lambda=0$), as well as with the SU(2) symmetric interaction
($\lambda=1$). For the density-density calculations, we employ the
same parameters as those used in Ref.~\onlinecite{Kunes2016}: $U=4$,
$J=1$, $U'=U-2J$, $\Delta_{\rm CF}=3.4$, $t_a=0.4118$, $t_b=-0.1882$, $V=0.05$,
while for the SU(2) symmetric calculations, we set $V=0$ or $V =
0.04$.

The model is investigated in the DMFT approximation, where the lattice
model is mapped onto the Anderson impurity model that interacts with
an effective bath~\cite{Georges1996}. The auxiliary impurity problem
is solved numerically using the continuous-time quantum Monte-Carlo
(CT-QMC) algorithm in the hybridization expansion
formalism\cite{Werner2006a}. The hybridization function which
describes the interaction between the impurity and the bath states is
defined by
\begin{equation}
  F_{\gamma\gamma'}(\tau)=\frac{1}{\beta}
  \sum_{n=-\infty}^{\infty} e^{-i\omega_n\tau} F_{\gamma\gamma'}(i\omega),
\end{equation}
where $\tau$ is the imaginary time, and $F_{\gamma\gamma'}(i\omega)$ is given by
\begin{equation}
  F_{\gamma\gamma'}(i\omega)=
  \sum_{k} \frac{V^*_{\gamma k}V_{k\gamma'}}{i\omega_n-\epsilon_k}.
\end{equation}

Here, $\omega_n=(2n+1)\pi/\beta$ ($n$: integer) is the Matsubara
frequency and $\beta$ is the inverse temperature. The index $\gamma$
represents both the orbital and spin degrees of freedom of the impurity, e.g., $\gamma=\{a,\uparrow \}$. The index $k$ labels the bath state $k$ with energy
$\epsilon_k$. In model studies using DMFT,
$F_{\gamma\gamma'}(\tau)$ is often considered (due to symmetry) or approximated to be diagonal and real.
In
the present model, however, the finite off-diagonal component of 
$F_{\gamma\gamma'}$ represents the Weiss field of the excitonic phase
and is, therefore, central to our study.


\subsection{Excitonic order parameter and other observables}

In the present model, the exciton condensate can be characterized by
inspecting the site-independent orbital-off-diagonal elements of the
local occupation matrix $\Delta_{\alpha\beta}\equiv \expval{a_{\bm{i}
    \alpha}^{\dagger}b_{\bm{i}\beta}^{\phantom\dagger}}$. In the
normal state, $\Delta$ is proportional to a unit matrix. In the
condensate, a spin triplet component appears, that can be described by
a complex vector order parameter
\begin{equation}
  \phi^\gamma \equiv \trace( {\sigma^\gamma}^*\Delta) = \sum_{\alpha\beta} \sigma^\gamma_{\alpha\beta}
  \expval{a_{\bm{i} \alpha}^{\dagger}
    b_{\bm{i}\beta}^{\phantom\dagger}},
\end{equation}
where $\sigma^{\gamma}$ ($\gamma=x,y,z$) denotes the Pauli
matrices. In general $\bphi=\bR+i\bI$, where the real vectors
$\bR$ and $\bI$ transform according to SO(3) group
under spin rotations and as $\tau: \bphi\rightarrow -\bphi^*$
under time reversal. The complex nature of $\bphi$ allows various
excitonic phases that can be distinguished by
\endnote{We use the notation: $\bm{u}\cdot\bm{v} \equiv
  u_xv_x+u_yv_y+u_zv_z$ and ${\|\bm{v}\|^2 \equiv \bm{v}^*\cdot\bm{v}}$.}
\begin{equation}
  \| \bphi^*\times\bphi \|^2=(\bphi^*\cdot\bphi)^2 -|\bphi\cdot\bphi|^2.
\end{equation}
For the phases with ${\| \bphi^* \times \bphi \| = 0}$ the name
polar exciton condensate~\cite{ho98} (PEC) or unitary
phase~\cite{vollhardt1990superfluid} is used. The order parameter in
this case has the form ${\bphi= e^{i\theta}\bm{x}}$, with $\bm{x}$ a
real vector, and thus the phase has a residual uniaxial symmetry. If
$\bm{\phi}=i\bI$ the time reversal symmetry is preserved. 
Halperin and Rice~\cite{HALPERIN1968} introduced the names
spin-current-density wave (SCDW) condensate for this case 
and spin-density-wave (SDW) condensate for $\bm{\phi}=\bR$. 
The SDW phase exhibits a spin density distribution polarized along
$\bR$. The SCDW phase possesses a pattern of spin current polarized
along $\bI$.

A finite ${\|\bphi^ * \times \bphi \|}$ implies that
$\bR\nparallel\bI$ and so the condensate has no axial symmetry. The
most prominent feature of this phase is the appearance of a finite spin
moment ${\bM_\perp \propto i\bphi^ * \times
  \bphi}$\cite{balents00b,Kune2015} perpendicular to both $\bR$ and
$\bI$, which gives this phase its name ferromagnetic exciton
condensate (FMEC)~\cite{ho98}.

Besides the order parameter $\bphi$ and the local occupation matrix
we evaluate the reciprocal space spin texture
\begin{equation}
  \mk=\sum_{\alpha\beta}\boldsymbol{\sigma}_{\alpha\beta}
  \langle 
  a^\dagger_{\bk \alpha} a^{\phantom \dagger}_{\bk \beta} + 
  b^\dagger_{\bk \alpha} b^{\phantom \dagger}_{\bk \beta} 
  \rangle, 
\end{equation}
where $\bk$ is the reciprocal space vector and the $\bk$-indexed
operators are Fourier transforms of their local counterparts,
$a_{\bk}=\tfrac{1}{\sqrt{N}}\sum_{\bi} \exp^{-i\bk\cdot\bi}a_\bi$. A
finite $\bk$-odd contribution to $\mk$ may indicate the existence of a
net spin current, which we evaluate from
\begin{equation}
  \label{eq:spin_current}
  \begin{split}    
    J_{\nu}^\gamma = -2&\sum\limits_{\bk,\alpha\beta}
    \left(
      t_a \langle a_{\bk\alpha}^{\dagger}a_{\bk \beta}^{\phantom\dagger}\rangle
      +t_b \langle  b_{\bk\alpha}^{\dagger} b_{\bk\beta}^{\phantom\dagger}\rangle
    \right.
    \\ 
    &\left.
      + V_{+\nu}\langle
      a_{\bk\alpha}^{\dagger}b_{\bk\beta}^{\phantom\dagger}\rangle   +
      V_{-\nu}\langle
      b_{\bk\alpha}^{\dagger}a_{\bk\beta}^{\phantom\dagger}\rangle 
    \right)
    \sigma^\gamma_{\alpha \beta} \sin k_\nu.
  \end{split}
\end{equation}
The derivation can be found in Appendix~\ref{sec:app1}.

\section{Results and discussion}
\subsection{The $\mathbf{h}=0$ case}
\label{sec:h0}
First, we discuss the order parameter $\bphi$ and the net spin 
polarization $\bM$ in various excitonic phases in the absence of
external field. Although there is no continuum spin density defined in
our lattice model, one can say whether a continuum spin density exists
or identically vanishes assuming our model is built on real Wannier
orbitals. Generally, a finite $\bphi$ gives rise to a spin density
proportional (and parallel) to $\bR$ and spin current density
proportional to $\bI$~\cite{HALPERIN1968,Kune2015}.

We start with the density-density form of the
on-site interaction ($\lambda=0$), which effectively introduces 
an easy-axis magneto-crystalline anisotropy and restricts $\bphi$ to
the hard ($xy$) plane. Later we present results obtained with 
the rotationally invariant interaction and show that they exhibit the
same qualitative behavior. The density-density interaction allows
comparison with our previous work~\cite{Kunes2016} and takes advantage
of the faster computational algorithm as well as higher transition
temperatures. The present results were obtained with two independent
implementations of the complex hybridization in the CT-QMC
algorithm~\cite{Hoshino2016,hariki15}.

\begin{figure}
  \centering
  \includegraphics[width=\columnwidth]{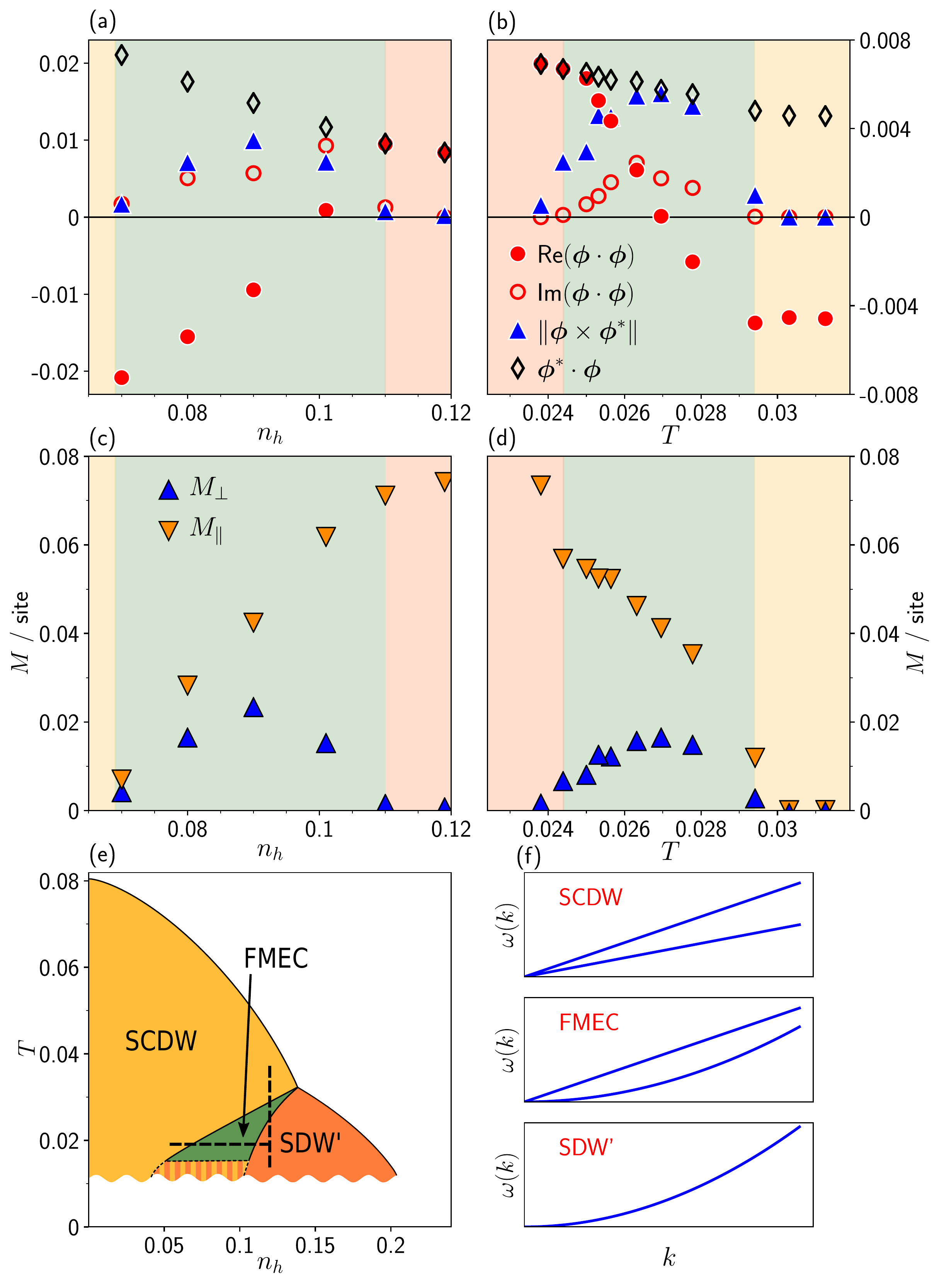}
  \caption{\label{fig:cplx_phi_combined} 
    Evolution of the order
    parameter $\bphi$ and the net spin moment $\bM$ along the constant
    doping (a,c) and constant temperature (b,d) cuts
    of the phase diagram (e) of the model with $s$-wave cross-hopping  
    pattern and density-density interaction. (f) A cartoon
    view of the Nambu-Goldstone modes of the SU(2) symmetric model.}
\end{figure}

\subsubsection{$s$-wave cross-hopping}
The $s$-wave cross-hopping is distinguished from the
other hopping patterns by a finite and real expectation value of
$\phi_0 \equiv \langle a^{\dagger}_{i\uparrow}
b^{\phantom\dagger}_{i\downarrow}+a^{\dagger}_{i\uparrow}
b^{\phantom\dagger}_{i\downarrow}\rangle$. This can be
viewed as a spin-singlet component of the exciton condensate
generated by a source field present in the Hamiltonian. Exciton 
condensates with finite singlet and triplet components
were shown~\cite{volkov73,volkov75,balents00b,Bascones2002} 
to host a ferromagnetic polarization with components 
{$\bM_{\perp} \propto i \bphi^* \times \bphi$},
{$ \bM_{\parallel} \propto \phi_0^* \bphi + \phi_0\bphi^*$}.
The same spin polarization pattern may be expected here.

In Fig.~\ref{fig:cplx_phi_combined} we show the phase diagram 
as a function of temperature and hole doping (relative to the
half-filling of two electrons per atom) and summarize
the evolution of the order parameter $\bphi$ and of the net spin
moment $\bM$ along two cuts crossing the SCDW, FMEC and SDW' phases.

\begin{figure}[H]
  \centering
  \includegraphics[width=\columnwidth]{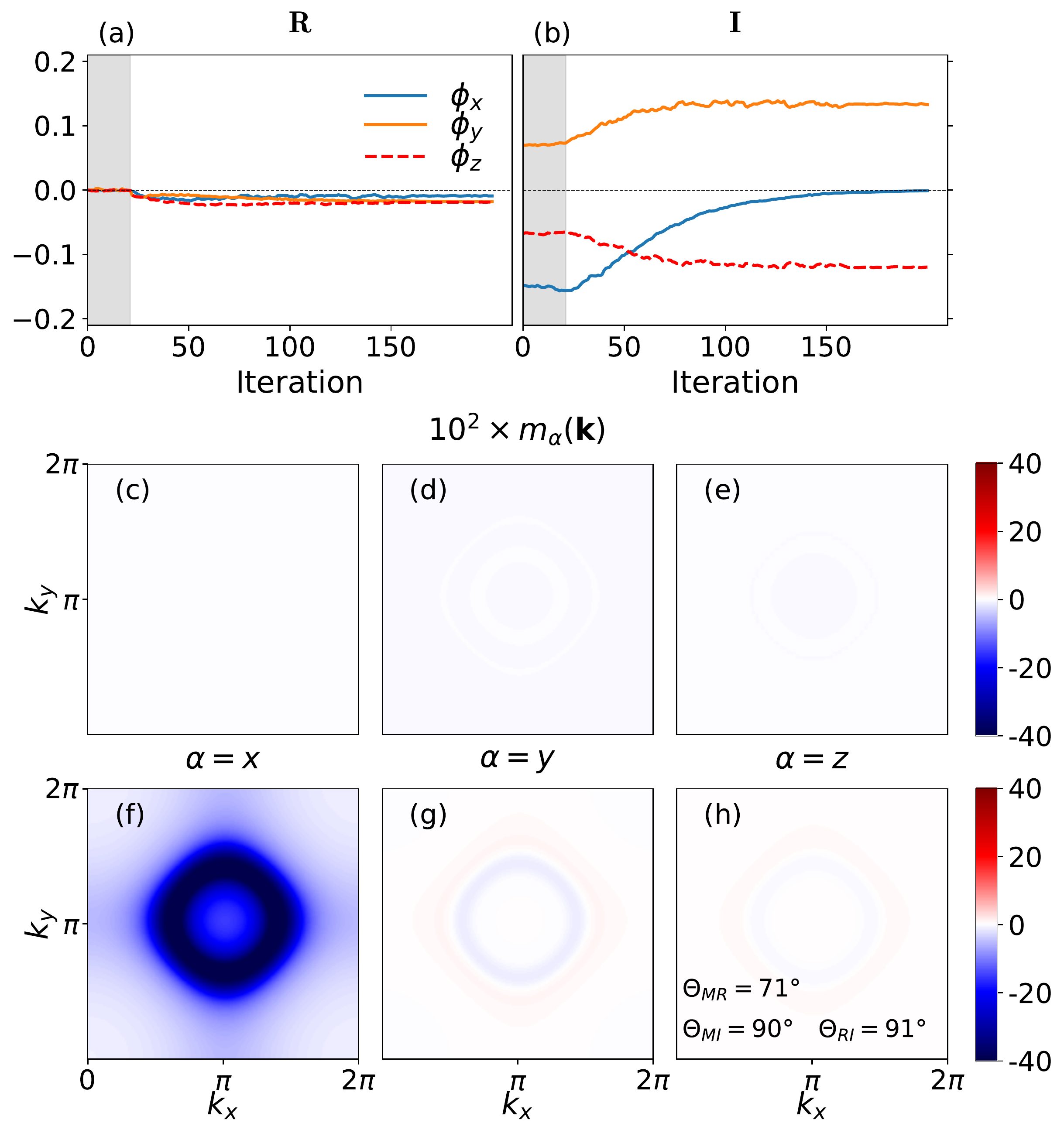}
  \caption{\label{fig:su2_rotation_scdw} 
  The SCDW phase of the SU(2) symmetric model with $s$-wave
  cross-hopping with and without an external magnetic field. (a,b)
  Convergence of the order parameter $\bphi$. The grey area marks the
  converged $\bh=0$ solution with arbitrary $\bphi$ orientation. The
  white area shows the convergence after a field along the $x$-axis is
  turned on. Panels (c-e) show the spin texture $\mk$ in the initial
  $\bh=0$ state. Panels (f-h) show $\mk$ in the converged solution
  with finite field. In panels (e) and (h) we show the angles  
  $\Theta_{MR}$, $\Theta_{MI}$ and $\Theta_{RI}$ between the vectors
  $\bM$, $\bR$ and $\bI$ (omitted if corresponding vectors vanish) in zero
  and finite field, respectively.
  The calculations were performed for $h_x=-0.006$,
  $n_h=0.03$, ${T=0.0125}$, $V = 0.04$.}
\end{figure}

In the SCDW phase $\bphi=i\bI$, implying 
${\re(\bphi \cdot\bphi) <0}$ and ${\im (\bphi\cdot\bphi) =0}$. The 
net moment ${\bM=0}$. The state is time-reversal invariant and thus
the continuous spin density vanishes as well. The spin currents
present in this state do not give rise to any spin texture,
$\mk=0$ (Fig.~\ref{fig:su2_rotation_scdw}c-e). In the SU(2) symmetric
model (Fig.~\ref{fig:su2_rotation_scdw}) there are two broken
generators of the SU(2) symmetry with vanishing expectation value of
their commutator (no net moment). This implies two linear Goldstone
modes~\cite{watanabe2012}.

In the SDW' phase $\bphi=\bR$, implying ${\re (\bphi \cdot \bphi) > 0}$, 
${\im (\bphi\cdot\bphi) =0}$. There is a net spin moment 
$\bM$ parallel to $\bR$ ($\bM_{\perp}=0$). 
There is naturally a finite spin texture 
$\mk$ parallel to $\bR$ (Fig.~\ref{fig:su2_rotation_sdw_prime}c-e), but
no texture in the perpendicular direction.
In the SU(2) symmetric case Fig.~\ref{fig:su2_rotation_sdw_prime} 
there are two broken generators of the SU(2) symmetry with finite 
expectation value of their commutator (the same as in a normal ferromagnet)
implying a single quadratic Goldstone mode~\cite{watanabe2012}.

At low temperatures the transition between the SCDW and SDW' phase
is of the first order. A continuous transition that we find at higher
temperatures and study here 
can possibly proceed via an intermediate polar phase or
an  FMEC phase. The latter is actually realized.
As the transition advances, $\bR$ and $\bI$ remain 
approximately perpendicular while changing their magnitudes
\endnote{ 
$\cos \Theta_{RI} = \tfrac{1}{2}\tfrac{\im(\bphi \cdot
\bphi)} {(\bphi^*\cdot\bphi)^2 - (\re (\bphi \cdot
\bphi))^2}$}. 
In the FMEC phase both $\bM_{\parallel}$ and
$\bM_{\perp}$ are finite and the net magnetization $\bM$ lies at a
general angle to both $\bR$ and $\bI$. The spin texture $\mk$
is found in both directions, parallel and perpendicular to $\bM$,
but with different structure (Fig.~\ref{fig:su2_rotation_swave_fmec}c-e).
In the SU(2) symmetric case, see Fig.~\ref{fig:su2_rotation_swave_fmec}, 
all three generators of the SU(2) symmetry are broken. The
finite expectation value of their commutator(s) ($\bM$) implies one quadratic
and one linear Goldstone mode~\cite{watanabe2012}.

\subsubsection{p,d-wave cross-hopping}\label{sec:pwave}
\begin{figure}
  \centering
  \includegraphics[width=\columnwidth]{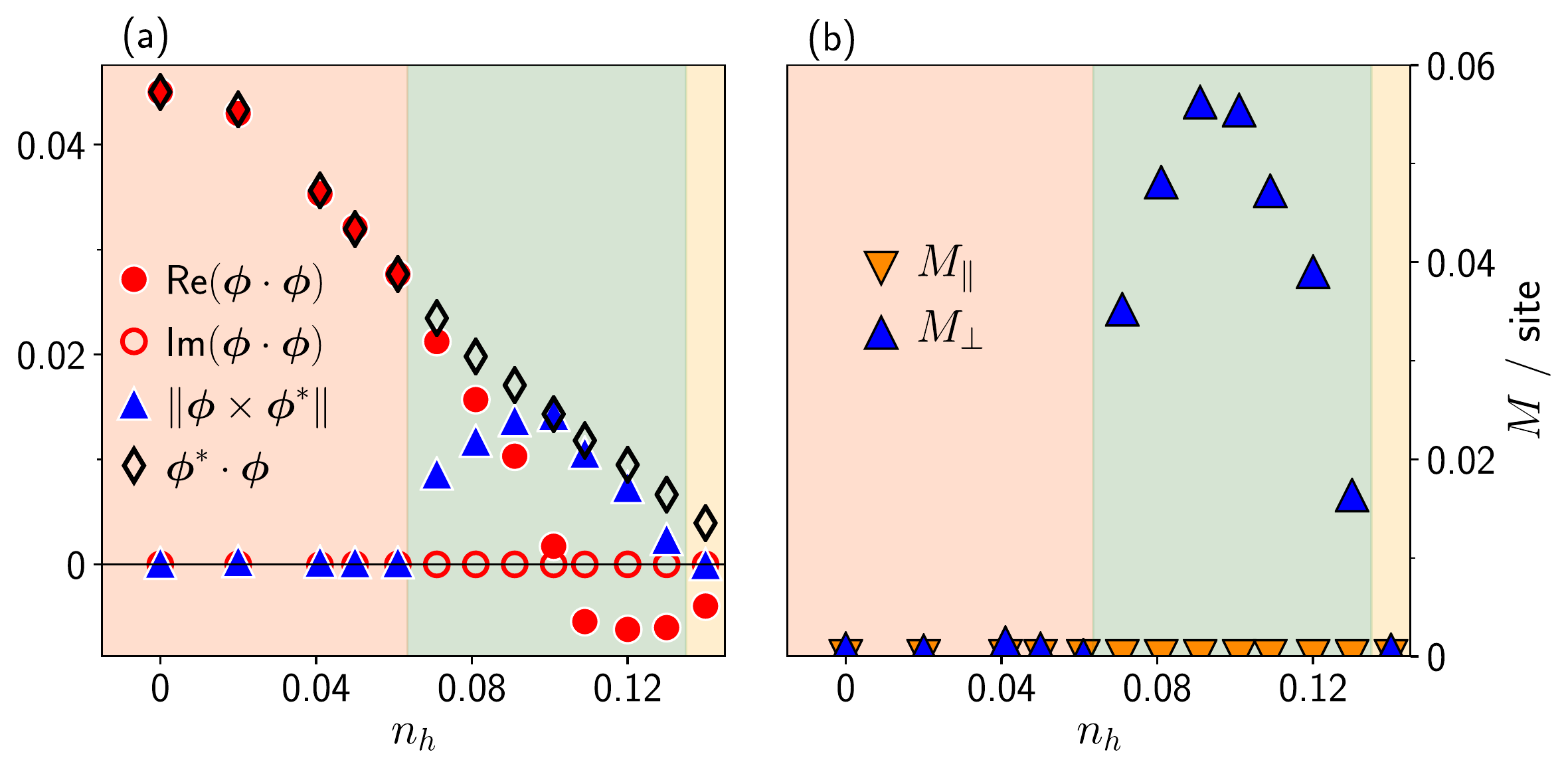}
  \caption{\label{fig:cplx_phi_combined_pwave} Evolution of the order
    parameter $\bphi$ (a) and the net spin moment $\bM$ (b) at
    a constant temperature $T=0.017$ for
    the model with $p$-wave cross-hopping pattern and 
    density-density interaction~\cite{Kunes2016}.}
\end{figure}

In the models with higher-$l$ cross-hopping pattern,
the local expectation value $\phi_0$ is zero in both the normal 
and the ordered phases, and polar phases have vanishing
ordered moments. In Fig.~\ref{fig:cplx_phi_combined_pwave} we show the
evolution of $\bphi$ along a cut in the phase diagram. Note that the
SDW and SCDW phases are exchanged relative to the $s$-wave case,
due to the opposite sign of
$V_{+\nu}V_{-\nu}$~\cite{Kunes2016,kunes14b}.

The SDW phase is characterized by $\bphi=\bR$, implying 
${\re (\bphi \cdot \bphi) > 0}$, ${\im (\bphi \cdot \bphi) =0}$.
Both $\bM$ and $\mk$ are absent (Fig.~\ref{fig:su2_rotation_sdw}c-e). 
There is, however, a local (anti-ferromagnetic) distribution of 
continuum spin density around each lattice site polarized along $\bR$.
The Goldstone spectrum consists of two linear modes.

The SCDW' phase is characterized by $\bphi=i\bI$, implying
${\re (\bphi \cdot \bphi) < 0}$, ${\im (\bphi \cdot \bphi) =0}$. 
The continuum spin density vanishes everywhere and the state is
time-reversal invariant. Unlike the SCDW phase in the $s$-wave case,
there is a finite spin texture $\mk$ with $p$-wave
symmetry~\cite{Kunes2016} (Fig.~\ref{fig:su2_rotation_xy}c-e), which can
be viewed as a $\bk$-space anti-ferromagnet. The Goldstone spectrum
again consists of two linear modes.

Similar to the $s$-wave case, the transition between the SDW and SCDW' phases
is of first order at low temperatures, and continuous via an intermediate
FMEC phase at higher temperatures. Unlike the $s$-wave case,
$\bM_{\parallel}=0$ and $\im (\bphi\cdot\bphi)$=0 along the path. The
vectors $\bM$, $\bR$ and $\bI$ thus remain mutually orthogonal along 
the whole path through the FMEC phase. The transition
proceeds by shrinking of $\bR$ accompanied by growth of $\bI$. 
There is a $p$-wave spin texture $\mk$ in the $\bR$-$\bI$ plane
and an $s$-wave texture for perpendicular polarization,
Fig.~\ref{fig:su2_rotation_xy_d06}c-e. The Goldstone spectrum is the same
as in the FMEC phase of the $s$-wave model.

The model with $d$-wave cross-hopping is expected to show a 
behavior similar to the $p$-wave one, i.e. $\bM_\parallel=0$ and $\bR \perp
\bI$. The roles of the SDW and SCDW phases are exchanged due to the
same sign of $V_{+\nu}V_{-\nu}$ as in the $s$-wave case. The spin
texture $\mk$ exhibits a $d$-wave symmetry in the SDW' phase. We have
not performed a systematic study, but confirmed this conclusion by
inspecting a selected point in each of the FMEC and SDW' phases.

\begin{figure}
  \centering
  \includegraphics[width=\columnwidth]{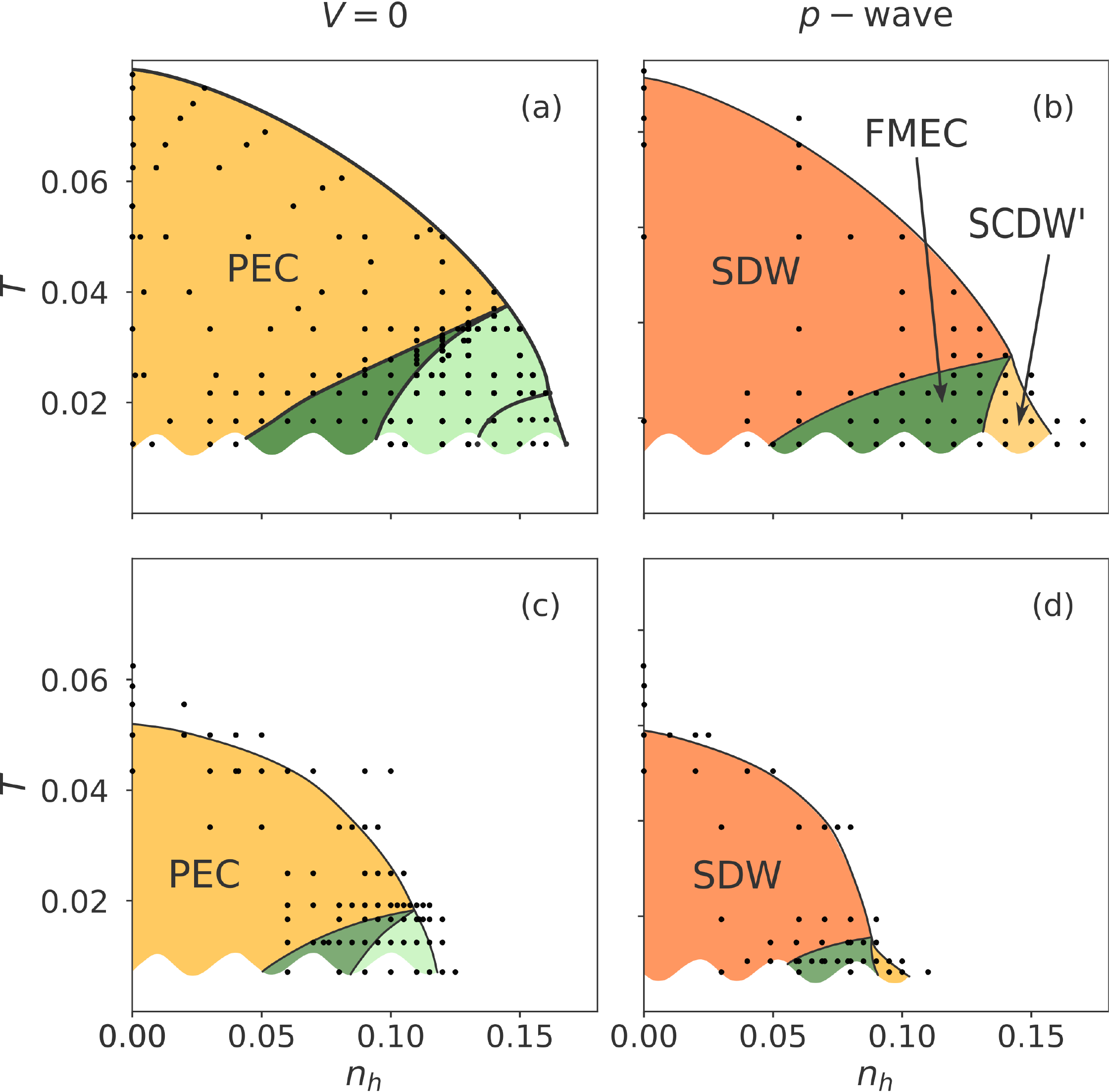}
  \caption{\label{fig:su2_phase_diag}
    Comparison of models with density-density (a,b) and SU(2) symmetric
    (c,d) interaction. Panels (a,c) correspond to zero cross-hopping.
    Panels (b,d) were obtained with a $p$-wave cross-hopping
    pattern. The dots mark the points for which the calculations were
    actually performed.}
\end{figure}

\subsubsection{Rotationally invariant interaction}
Fig.~\ref{fig:su2_phase_diag} illustrates
the modification of the phase boundaries due to the spin-flip
term. Panels (a) and (c) show the $V_{\nu}=0$ case, while panels (b) and
(d) correspond to the $p$-wave cross-hopping pattern. The results for the
SU(2) symmetric model are qualitatively similar to the density-density
case, but the extent of the excitonic phase is reduced. This can be
traced back to the higher local degeneracy of the Heisenberg HS
state, which favors the normal phase.

\begin{figure}
  \centering
  \includegraphics[width=\columnwidth]{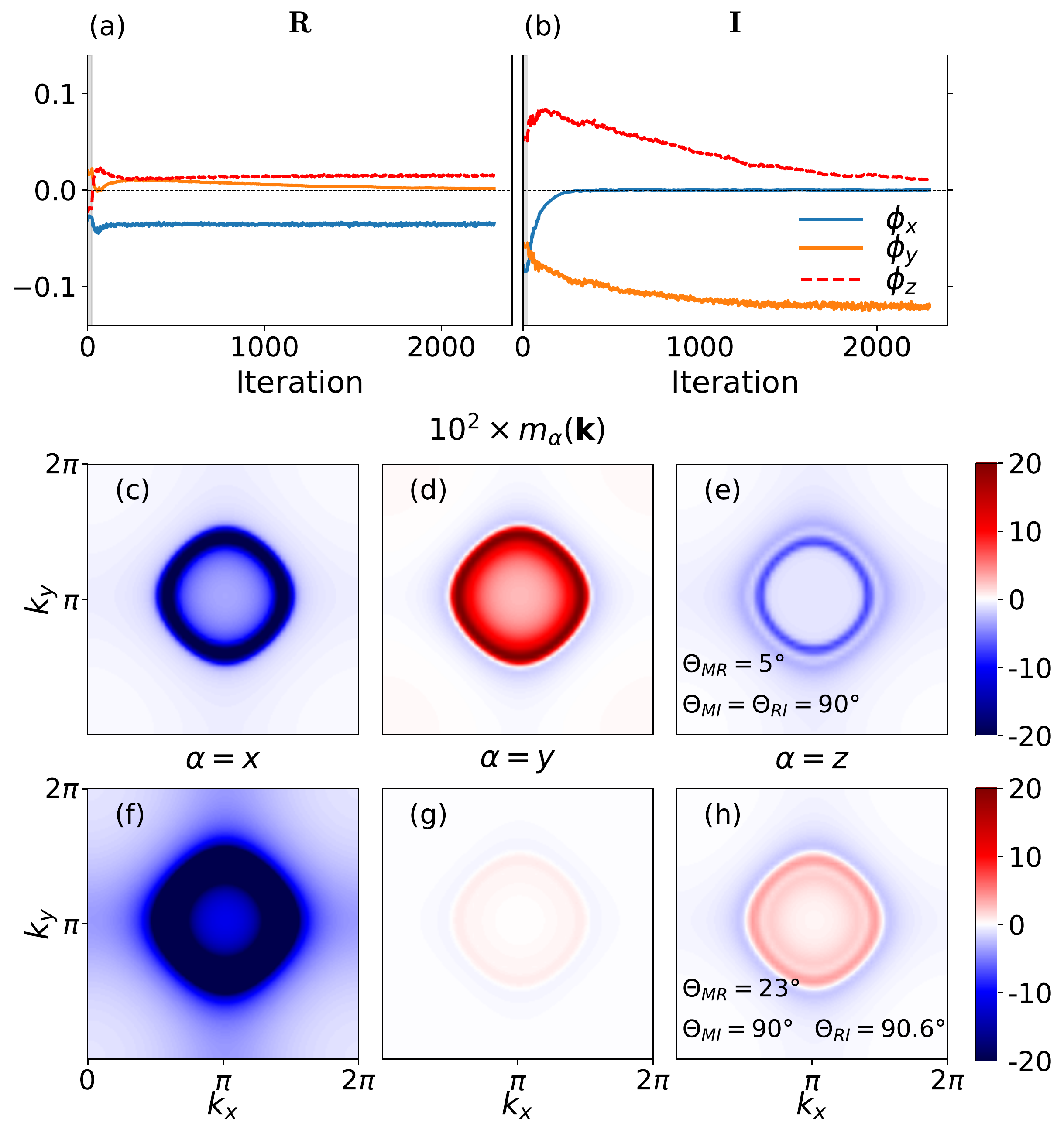}
  \caption{\label{fig:su2_rotation_swave_fmec} The same as
    Fig.~\ref{fig:su2_rotation_scdw} for the FMEC phase of the SU(2)
    symmetric model with $s$-wave cross-hopping. The calculations were
    performed for $h_x=-0.006$, $n_h=0.07$,
    $T=0.007$, $V = 0.04$. Note that calculations are practically converged 
    after $\sim$800 iterations. The evolution of $\bphi$ after this point
    represents mainly rotation around $\bh$, i.e., a symmetry transformation
    of the practically converged solution.}
\end{figure}

\begin{figure}
  \centering
  \includegraphics[width=\columnwidth]{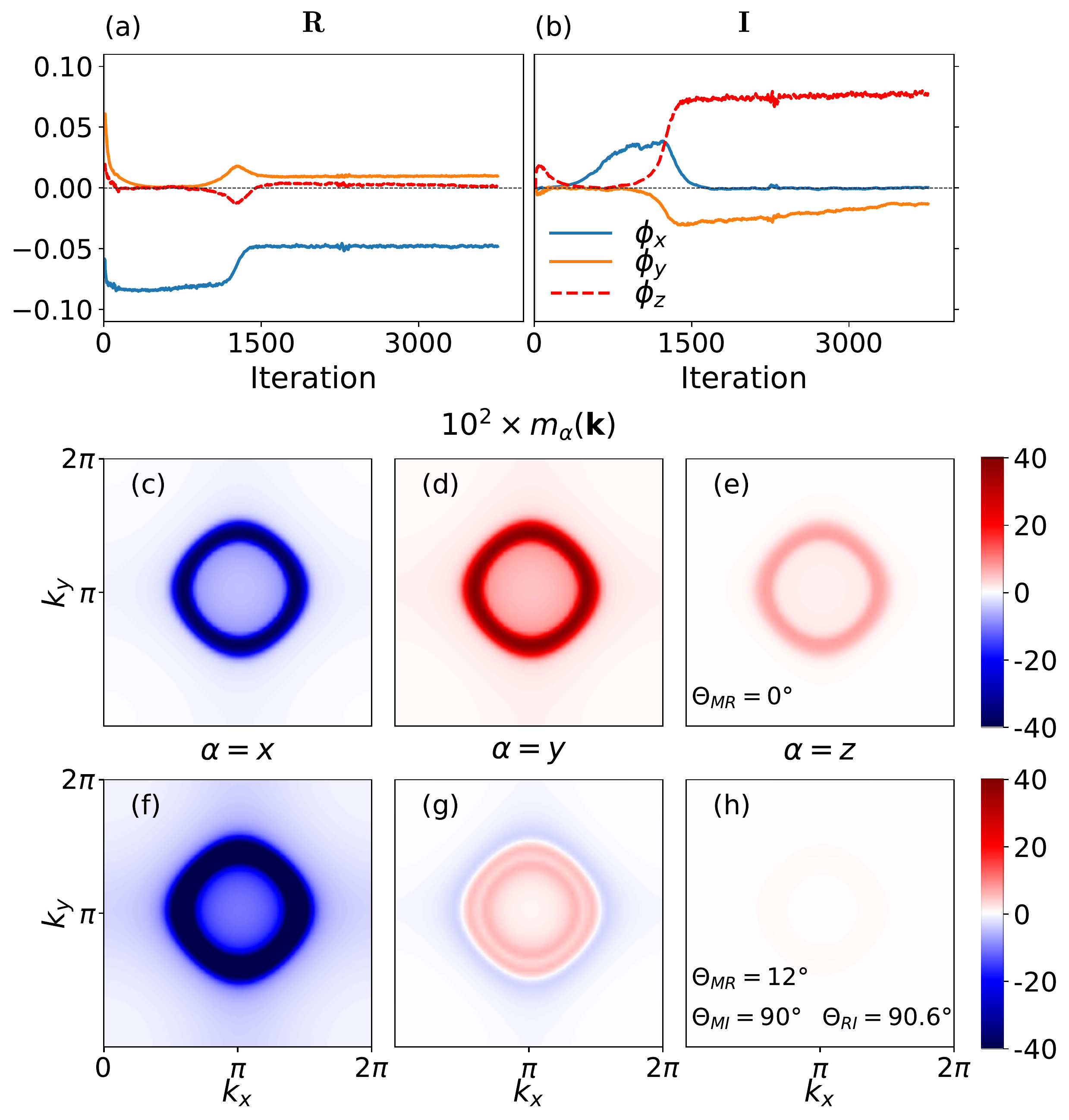}
  \caption{\label{fig:su2_rotation_sdw_prime}The same as
    Fig.~\ref{fig:su2_rotation_scdw} for the SDW' phase of the SU(2)
    symmetric model with $s$-wave cross-hopping. The calculations were
    performed for $h_x=-0.006$, $n_h=0.08$,
    $T=0.0125$, $V = 0.04$.}
\end{figure}

\subsection{Spin-triplet condensate in external field}
\label{sec:h}
Next, we study the condensate in small magnetic
(Zeeman) fields. In particular, we want to investigate the
orientation of the order parameter $\bphi$ with respect to the net
moment $\bM$ (parallel to the external field). To this end we use the
SU(2) symmetric interaction ($\lambda=1$). We start from a converged
$\bh=0$ result with $\bphi$ pointing in a general direction. Then a
magnetic field $\mathbf{h}$ pointing along the $x$-axis is applied and
convergence to the new equilibrium monitored. The field magnitude is
chosen to be smaller than the excitonic Weiss field, estimated as the
value of the high-frequency limit of the off-diagonal self-energy, but
large enough to achieve reasonably fast convergence of the DMFT
iterative procedure. For each excitonic phase we show the convergence
of $\bphi$, the spin texture $\mk$ in zero and finite $\mathbf{h}$ and
the angles $\Theta_{MR}$, $\Theta_{MI}$ and $\Theta_{RI}$ between the vectors
$\bM$, $\bR$ and $\bI$.

\subsubsection{$s$-wave cross-hopping}
Starting from a small doping we first visit the SCDW phase, see
Fig.~\ref{fig:su2_rotation_scdw}. As with the density-density
interaction at $\bh=0$, this phase is characterized by
${\bR = \bM = \mk = 0}$ and a pattern of local spin currents
polarized parallel to $\bI$, as discussed in the previous section. 
The dominant effect of the external 
field $\bh$ is to rotate $\bI$ perpendicular to $\bh$. This
behavior, reminiscent of an antiferromagnet, will be observed also in
other cases. A small $\bR$ component is induced, which is
approximately perpendicular to $\bh$ and $\bI$.

The FMEC state obtained at higher doping carries a finite net moment
$\bM$, see Fig.\ref{fig:su2_rotation_swave_fmec}c-e. 
The main effect of the external field $\bh$ is to align 
$\bM$ along its direction. The finite component of $\bR$
perpendicular to $\bM$ gives rise to an $s$-wave spin texture that is
not parallel to $\bM$ and thus integrates to zero over the Brillouin
zone, see Fig.~\ref{fig:su2_rotation_swave_fmec}g-h.

The SDW' phase also carries a finite spin polarization $\bM$,
Fig.~\ref{fig:su2_rotation_sdw_prime}c-e. While a vanishingly small $\bh$
would just rotate the ground state to align $\bM$ with $\bh$, the
finite field has a more profound effect. It gives rise to a
sizable $\bI$ and effectively induces a transition to an FMEC-like
state. It is interesting to point out that while in zero field the
SDW' phase has the same uniaxial symmetry as an ordinary ferromagnet,
this symmetry is lost in a finite field. It is instructive to inspect
the convergence of the iterative procedure, after $\bh$ is turned
on. First, the system remains in a unstable SDW'-like state ($\bI
\approx 0$ and $\bR\parallel\bh$) to eventually settle in an FMEC-like
state. Although the convergence does not represent any real dynamics of
the system, it suggests the existence of a metastable SDW'-like phase.

\subsubsection{$p$-wave cross-hopping}
\begin{figure}
  \centering
  \includegraphics[width=\columnwidth]{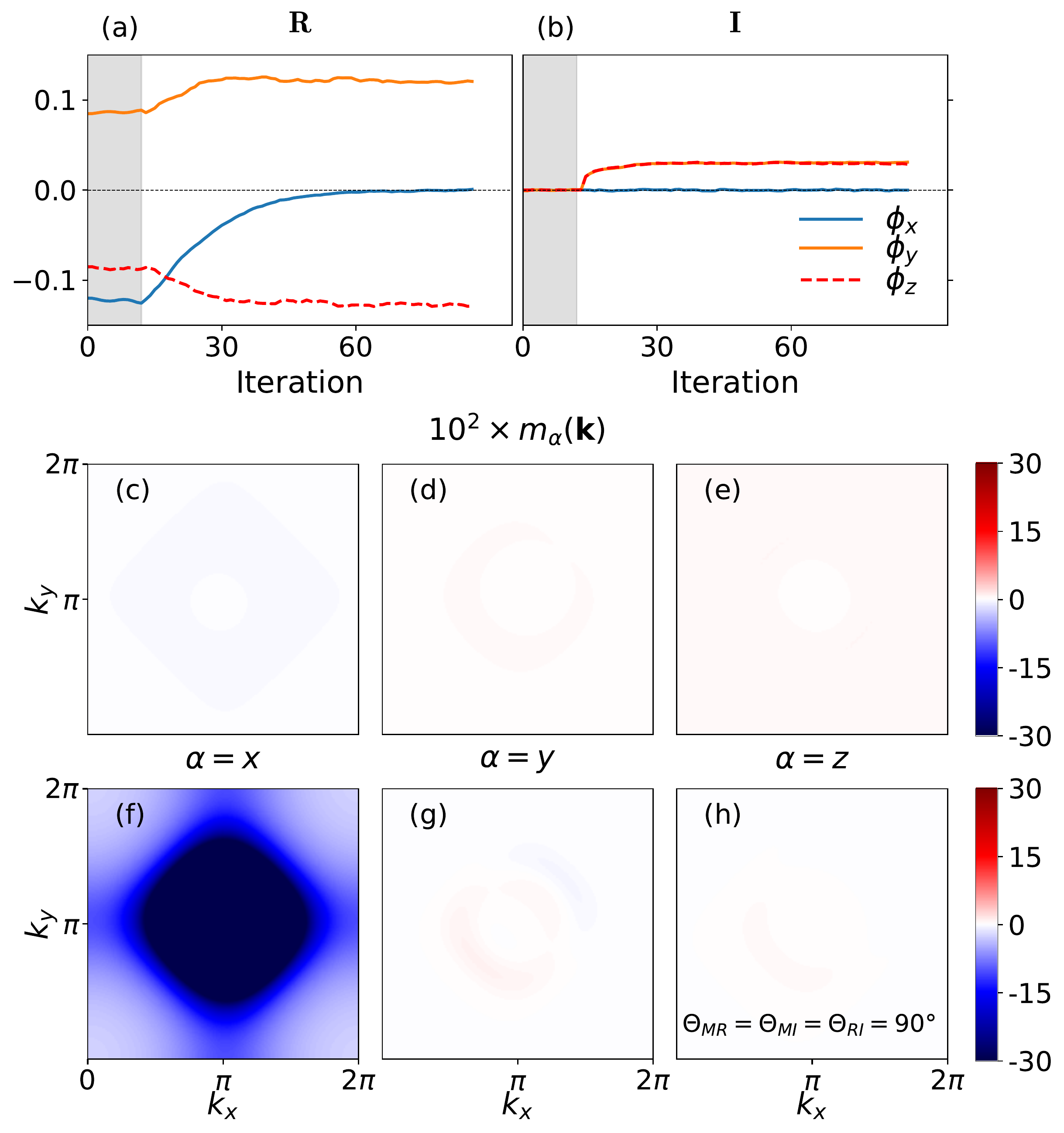}
  \caption{\label{fig:su2_rotation_sdw}The same as
    Fig.~\ref{fig:su2_rotation_scdw} for the SDW phase of the SU(2)
    symmetric model with $p$-wave cross-hopping. The calculations were
    performed for ${h_x=-0.01}$, ${n_h=0.03}$,
    ${T=0.033}$, ${V = 0.04}$.}
\end{figure}

At small doping and $\bh=0$ the system is in the SDW phase
characterized by ${\bI=\bM=\mk=0}$ and finite intra-atomic (collinear
antiferromagnetic) spin polarization parallel to $\bR$. In the
external field $\bh$, $\bR$ turns perpendicular to $\bh$, see
Fig.~\ref{fig:su2_rotation_sdw}. A small $\bI$ component perpendicular
to $\bR$ and $\bh$ is induced together with a small net moment.

\begin{figure}
  \centering
  \includegraphics[width=\columnwidth]{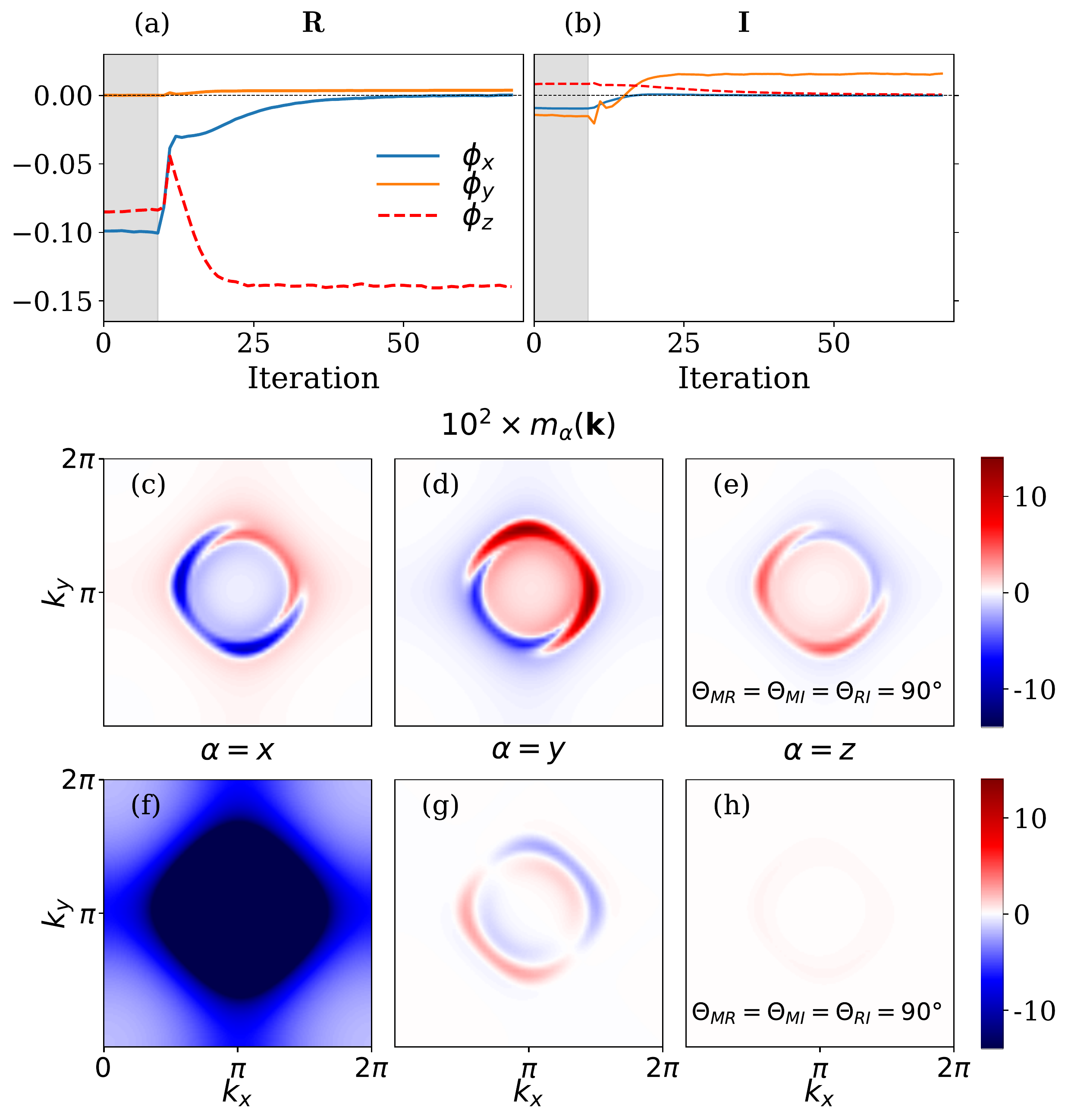}
  \caption{\label{fig:su2_rotation_xy_d06}The same as
    Fig.~\ref{fig:su2_rotation_scdw} for the FMEC phase of the SU(2)
    symmetric model with $p$-wave cross-hopping. The calculations were
    performed for ${h_x=-0.01}$, ${n_h=0.6}$,
    ${T=0.007}$, ${V = 0.04}$.}
\end{figure}

Applying finite $\bh$ in the FMEC phase aligns the spontaneous polarization
$\bM$ with the external field $\bh$ as expected, while the mutual
orthogonality of $\bM$, $\bR$ and $\bI$ is preserved. The polarization
of the $p$-wave spin texture thus remains perpendicular to $\bh$, see
Fig.~\ref{fig:su2_rotation_xy_d06}.

\begin{figure}
  \centering
  \includegraphics[width=\columnwidth]{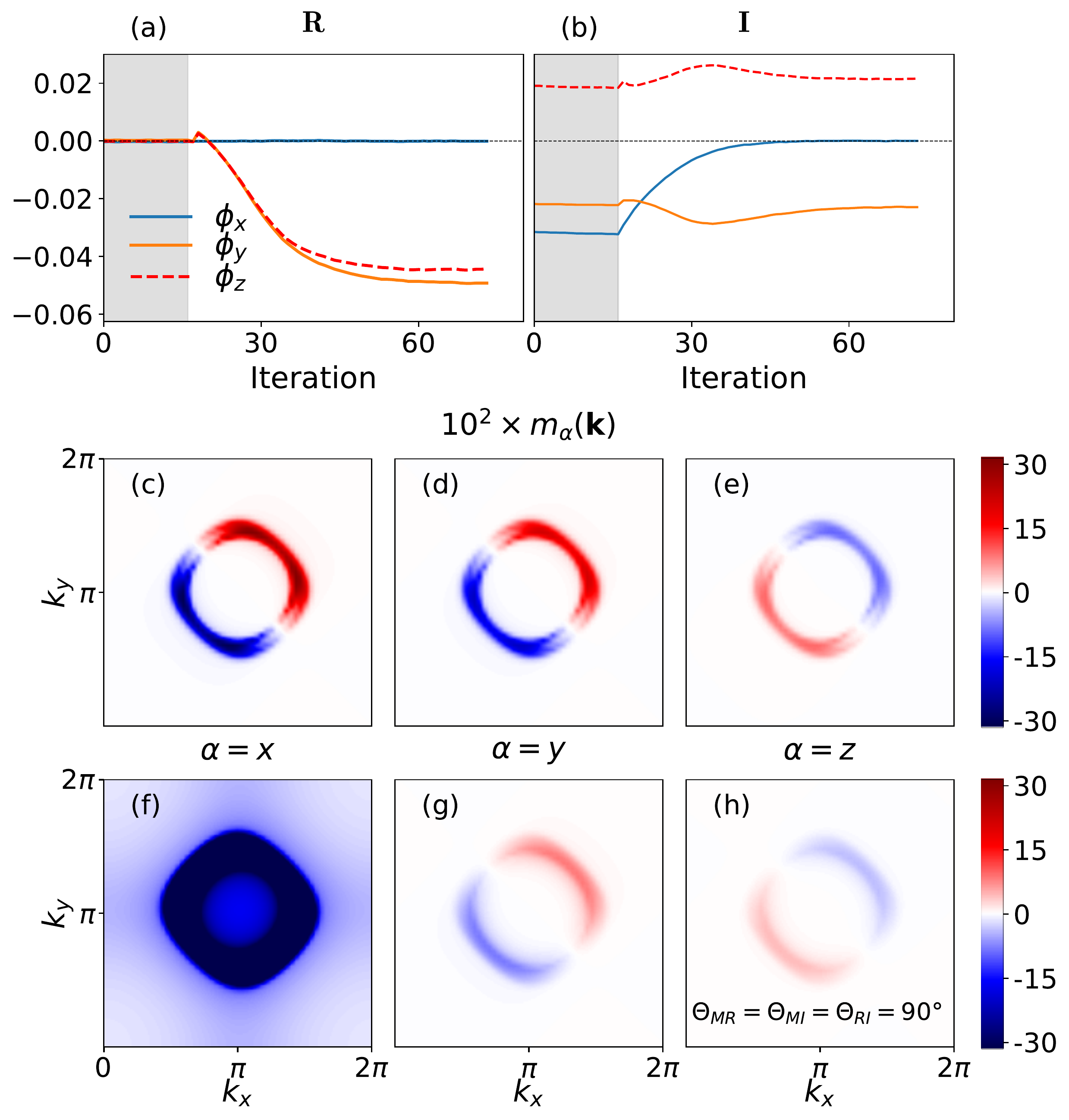}
  \caption{\label{fig:su2_rotation_xy}The same as
    Fig.~\ref{fig:su2_rotation_scdw} for the SCDW' phase of the SU(2)
    symmetric model with $p$-wave cross-hopping. The calculations were
    performed for $h_x=-0.01$, $n_h=0.095$,
    ${T=0.007}$, $V = 0.04$.}
\end{figure}

Finally, the SCDW' phase at $\bh=0$ is invariant with respect to time
reversal and thus carries no spin polarization. Nevertheless, the
spin-rotational symmetry is broken, as demonstrated by the presence of
the spin texture $\mk$. A finite external field $\bh$ generates
a state similar to the FMEC case with the spin texture polarized
perpendicular to $\bM$, see Fig.~\ref{fig:su2_rotation_xy}.

In fact, with finite $\bh$, all the excitonic phases become equivalent
to the FMEC phase, although obvious quantitative differences remain
for the case of moderate $\bh$ discussed here. We point out that there
is still a symmetry difference between the excitonic condensate and
the normal state in the presence of finite field, since the condensate
does not have the uniaxial symmetry of the normal state.

\subsection{Phenomenological model}
\label{sec:pheno}
The above numerical results paint a rather complex picture. In
order to understand them we introduce a phenomenological
Ginzburg-Landau type (GL) functional, which can be viewed as an
extension of the functional of Ref.~\onlinecite{kunes14c}. We assume
that the magnitude of the order parameter $\|\bphi\|^2$ is fixed by
the large kinetic energy of excitons and show only the smaller terms
that select the direction of $\bphi$. 
We start by considering an undoped system. The corresponding GL
functional reads
\begin{equation}
  \label{eq:gl1}
  E_0=\pm\alpha(\bR\cdot\bR-\bI\cdot\bI)+\beta(\bR\times\bI)\cdot(\bR\times\bI)
  -\bh\cdot(\bR\times\bI),
\end{equation}
with positive constants $\alpha$ and $\beta$. Here the first term describes
the effect of cross-hopping on the phase of the order parameter. The
plus sign applies to $s$- and $d$-wave cross-hopping, the minus sign
to $p$-wave
cross-hopping. $\bR\times\bI=-\tfrac{i}{2}\bphi^*\times\bphi$ is
proportional to the spin polarization of the condensate, so that the
second and third terms describe the inter-atomic antiferromagnetic
interaction and coupling to the external field, respectively. For
$\bh=0$, $\beta > 0$ implies  $\bR=0$ for $s$, $d$-wave cross-hopping
patterns, and $\bI=0$ for the $p$-wave pattern.

The application of a finite external field induces a
non-zero complementary component
\begin{equation*}
\bR = -\frac{\bh\times\bI} {2[C \pm \alpha + \beta\bI\cdot\bI]}, \quad
\bI = \frac{\bh\times\bR} {2[C \mp \alpha + \beta\bR\cdot\bR]},
\end{equation*}
where $C$ is a Lagrange multiplier fixing $\|\bphi\|^2$. This explains
the numerical observation of mutual orthogonality of $\bh$, $\bR$ and
$\bI$ in the undoped phases. It also justifies the use of
the density-density approximation with the field applied along the
$z$-axis, i.e. perpendicular to the condensate~\cite{Kunes2014b}.

Doping introduces additional terms to the functional. To proceed we
start from the generalized double-exchange
model~\cite{Kunes2016,chaloupka16}. We introduce terms that describe
the polarization of the doped carriers and its coupling to the
condensate
\begin{equation}
  \label{eq:gl2}
  E=E_0 + \gamma \bP \cdot (\bR \times \bI) - \bh \cdot \bP + \delta
  \bP \cdot \bR + \omega \bP \cdot\bP,
\end{equation}
where $\bP$ stands for the spin polarization of the doped
carriers. The second term describes the standard double-exchange
interaction between the local moments of the condensate and the
itinerant carriers. The third and fifth terms ($\omega>0$) describe the
polarizability of the doped carriers. The fourth term describes the
coupling between the condensate and the doped carriers due to the
finite cross-hopping. This term has a more complicated
$\bk$-dependent form~\cite{Kunes2016}, but to discuss
the response to a uniform field we keep only the part
containing $\bP$. The key observation is
that $\delta\neq 0$ for $s$-wave cross-hopping, while
$\delta=0$ for $p$- and $d$-wave cross-hopping.

The stationary values of $\bR$, $\bI$ and $\bP$ satisfy 
\begin{equation*}
\begin{split}
&\bR = \frac{\bI \times (\bh - \gamma \bP)  - \delta \bP} {2
    [ C \pm \alpha + \beta \bI \cdot \bI]},\quad
    \bI = \frac{\bR
    \times (\gamma \bP -\bh)}{2[C \mp \alpha + \beta \bR \cdot
    \bR]} \\
    &\bP = \frac{1}{2 \omega} (\bh - \gamma \bR \times
  \bI - \delta \bR).
  \end{split}
  \end{equation*}
This implies the orthogonality of $\bI$ to $\bP$,
$\bh$ and $\bR$, as observed in the numerical calculations. For $p$-
and $d$-wave cross-hopping, $\delta=0$ so that $\bR$ is
orthogonal to $\bh$ and $\bP$ (which are parallel in this
case) as well. Finite $\delta$ in the $s$-wave case leads
to a general angle between the coplanar vectors $\bR$, $\bP$ and
$\bh$. This behavior of the $s$-wave model reproduces the numerical
results only approximately. While $\bI\cdot\bh=0$ is fulfilled to our
numerical accuracy, we find small, but non-negligible, deviations from
$\bI\cdot\bR=0$, which must be due to effects beyond Eq.~\ref{eq:gl2}.

\subsection{Spontaneous spin current}
\label{sec:current}
The spin texture with $\mathbf{m}_{-\mathbf{k}}=-\mk$ in the SCDW' and
FMEC phases for $p$-wave cross-hopping,
Figs.~\ref{fig:su2_rotation_xy_d06}-\ref{fig:su2_rotation_xy},
may suggest that electrons moving in opposite directions
carry opposite spin polarization. Things are not so simple, since the
current (\ref{eq:spin_current}) depends on the group velocities, which
have opposite signs for $a$ and $b$ orbitals of the present model.

\begin{figure}[h]
  \centering
  \includegraphics[width=\columnwidth]{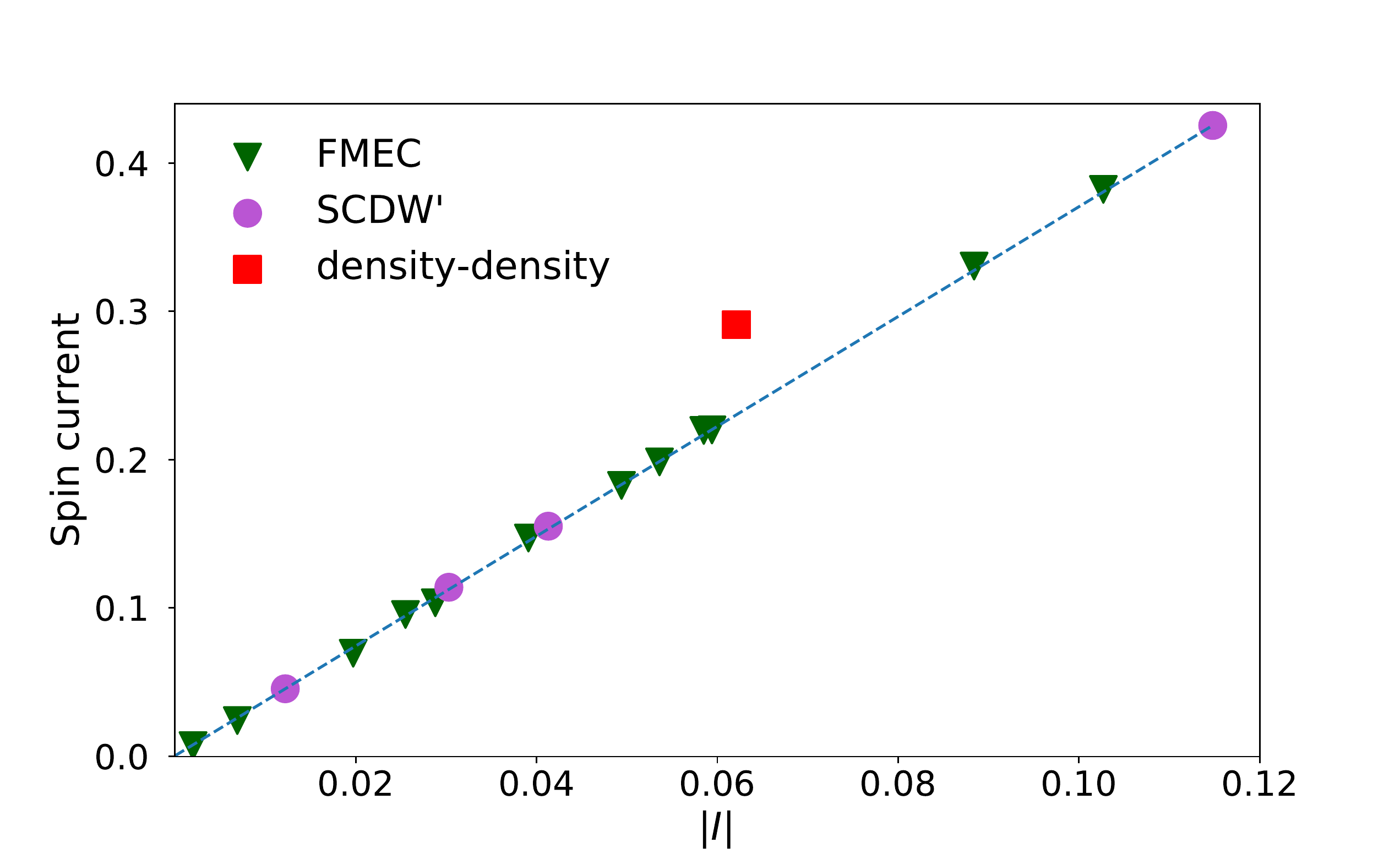}
  \caption{\label{fig:current_phase_diag}Magnitude of the spin current
    versus amplitude of the imaginary part of the order
    parameter, in two
    distinct phases using the SU(2) symmetric interaction (green
    triangles and magenta circles for FMEC and SCDW', respectively),
    and the density-density approximation (red square). The dashed
    line is a guide for the eye.}
\end{figure}

The calculated spin currents (\ref{eq:spin_current}) in the phases
with $p$-wave spin texture, marked by points in
Fig.~\ref{fig:su2_phase_diag}, are shown in
Fig.~\ref{fig:current_phase_diag}. We find a finite net spin current
polarized along, and scaling with $\bI$~\endnote{The spin polarization
  depends on $\bphi$ while the spatial orientation is given by the
  hopping pattern.}. This shows that DMFT violates the so-called Bloch
theorem~\cite{To1938,Ohashi1996}, which forbids spontaneous currents
of charges that are locally conserved by the interaction
Hamiltonian. In the Appendix~\ref{sec:app2} we sketch the proof for
the present model at $T=0$. A general proof for finite temperatures
can be found in Ref.~\onlinecite{Ohashi1996}.

Comparing Eq.~\ref{eq:spin_current} with the definition of the spin
texture $\mk$, it is clear that a vanishing of spin current does not
require that $\mk=0$. We assume that a spin texture exists also in the
exact ground state of the model, while the spin current is suppressed
by the momentum-dependence of the self-energy, absent in DMFT.
If so, a finite spin current may be obtained by breaking the balance
between the orbital contributions to (\ref{eq:spin_current}) in a
non-equilibrium state generated by an optical excitation.

\section{Conclusions}
We have studied the influence of an external magnetic field on the
excitonic condensate in the two-band Hubbard model. In all studied
phases the excitonic condensate breaks the uniaxial symmetry imposed
by the external field and the excitonic condensation thus  remains a
thermodynamic phase transition accompanied by the appearance of
gapless Goldstone modes.

There is a ubiquitous $\bh\cdot(\bR \times\bI)$ coupling between the
field and condensate, which generates perpendicular (to $\bh$)
components of the order parameter $\bphi$. As a result the staggered
spin density or spin current density polarization in the model with
$p$- and $d$-wave cross-hooping lie perpendicular to the external
field, analogous to the behavior of an Heisenberg antiferromagnet. For
$s$-wave cross-hopping, an additional linear coupling $\bh\cdot\bR$
exists giving rise to a more complicated behavior.

Finally, we have observed that a net spin current is spontaneously
generated in some excitonic phases with $p$-wave cross-hopping. We
attribute the violation of Bloch's no-go theorem to the local
self-energy approximation of the dynamical mean-field theory. We
propose that a net non-equilibrium spin or charge current may be
generated by a uniform orbital or spin-orbital selective excitation in
phase with the $p$-wave spin texture.

\section*{Acknowledgments}
This work has received funding from the European Research Council
(ERC) under the European Union’s Horizon 2020 research and innovation
program (Grant Agreement No. 646807-EXMAG). D.G. was supported in part
by the project MUNI/A/1310/2016. Access to computing and
storage facilities owned by parties and projects contributing to the
National Grid Infrastructure MetaCentrum provided under the programme
``Projects of Large Research, Development, and Innovations
Infrastructures'' (CESNET LM2015042), and the Vienna Scientific
Cluster (VSC) is greatly appreciated. This work was supported by the
Ministry of Education, Youth and Sports from the Large Infrastructures
for Research, Experimental Development and Innovations project
``IT4Innovations National Supercomputing Center – LM2015070''.

\appendix

\section{Expression for the spin current\label{sec:app1}}

We consider the model Hamiltonian (\ref{eq:2bhm}). The
local charge and spin operators read
\begin{equation}
  \begin{aligned}
    n_{\bi}=&s^0_{\bi}=\sum_{\sigma}
    \left(a_{\bi\sigma}^{\dagger}a^{\phantom\dagger}_{\bi\sigma} +
      b_{\bi\sigma}^{\dagger}b^{\phantom\dagger}_{\bi\sigma}\right)\\
    s^{\gamma}_{\bi}=&\sum_{\alpha,\beta} \sigma^{\gamma}_{\alpha\beta}
    \left(a_{\bi\alpha}^{\dagger}a^{\phantom\dagger}_{\bi\beta} +
      b_{\bi\alpha}^{\dagger} b^{\phantom\dagger}_{\bi\beta} \right),\\
  \end{aligned}
\end{equation}
where $\sigma^{\gamma}$ ($\gamma=0,x,y,z$) are the Pauli matrices. The
density operator commutes with the local part of the Hamiltonian
\begin{equation}
  [n_{\bi},H_{\text{loc}}]=[s^z_{\bi},H_{\text{loc}}]=0.\\
\end{equation}
For the SU(2) symmetric interaction, all the components
of the local spin operator commute with $H_{\text{loc}}$
\begin{equation}
  [s^{\gamma}_{\bi},H_{\text{loc}}]=0\quad\text{if}\quad\lambda=1\\
\end{equation}
We can define the current using the continuity equation, which takes
the form of Kirchhoff's first law
\begin{equation}
  \label{eq:kirchhoff}
  \partial_t s^{\gamma}_{\bi}=\sum_{\nu}
  \left(j^{\gamma}_{\bi-\benu\nu}-j^{\gamma}_{\bi\nu}\right),
\end{equation}
where $j^{\gamma}_{\bi\nu}$ is the current flowing on the bond
$\bi\rightarrow\bi+\benu$. The time derivative of the local
density operator can be evaluated using the equation of motion
\begin{equation}
  \partial_t
  s^{\gamma}_{\bi}=i[H,s^{\gamma}_{\bi}]=i\sum_{\nu}[T_{\nu}+T_{\nu}^{\dagger},s^{\gamma}_{\bi}].
\end{equation}
We distinguish between the 'right'- and the 'left'-hopping parts ($T$
and $T^{\dagger}$ respectively) of the kinetic energy for future
convenience. For the right-hopping part, we find
\begin{equation}
  \label{eq:commut1}
  \begin{split}
    &i[T_{\nu},s^{\gamma}_{\bi}]\\
    =&it_a
    \sum_{\alpha,\beta}\sigma^{\gamma}_{\alpha\beta}
    \left(a_{\bi+\benu\alpha}^{\dagger}a^{\phantom\dagger}_{\bi\beta} -
      a_{\bi\alpha}^{\dagger}a^{\phantom\dagger}_{\bi-\benu\beta}\right)\\
    +&it_b \sum_{\alpha,\beta}\sigma^{\gamma}_{\alpha\beta}\left(
      b_{\bi+\benu\alpha}^{\dagger}b^{\phantom\dagger}_{\bi\beta}-
      b_{\bi\alpha}^{\dagger}b^{\phantom\dagger}_{\bi-\benu\beta}\right)\\
    +&iV_{+\nu}
    \sum_{\alpha,\beta}\sigma^{\gamma}_{\alpha\beta}\left(
      a_{\bi+\benu\alpha}^{\dagger}b^{\phantom\dagger}_{\bi\beta}-
      a_{\bi\alpha}^{\dagger}b^{\phantom\dagger}_{\bi-\benu\beta}\right) \\
    +&iV_{-\nu}
    \sum_{\alpha,\beta}\sigma^{\gamma}_{\alpha\beta}\left(
      b_{\bi+\benu\alpha}^{\dagger}a^{\phantom\dagger}_{\bi\beta}-
      b_{\bi\alpha}^{\dagger}a^{\phantom\dagger}_{\bi-\benu\beta}\right).
  \end{split}
\end{equation}
The operator $s^{\gamma}_{\bi}$ is Hermitian, therefore
\begin{equation}
  \label{eq:commut2}
  [T_{\nu}^{\dagger},s^{\gamma}_{\bi}]=-[T_{\nu},s^{\gamma}_{\bi}]^{\dagger}.
\end{equation}
Combining Eqs.~\ref{eq:kirchhoff}, \ref{eq:commut1} and \ref{eq:commut2}, we get
\begin{equation}
  \begin{split}
    j^{\gamma}_{\bi\nu}=&-
    \sum_{\alpha,\beta}
    \left(i t_a \sigma^{\gamma}_{\alpha\beta}
      a_{\bi+\benu\alpha}^{\dagger}a^{\phantom\dagger}_{\bi\beta}
      + \text{H.c.} \right) \\
    &-\sum_{\alpha,\beta}
    \left(i t_b \sigma^{\gamma}_{\alpha\beta}
      b_{\bi+\benu\alpha}^{\dagger} b^{\phantom\dagger}_{\bi\beta}
      + \text{H.c.} \right) \\
    &-\sum_{\alpha,\beta}\left(
      i V_{+\nu} \sigma^{\gamma}_{\alpha\beta}
      a_{\bi+\benu\alpha}^{\dagger}b^{\phantom\dagger}_{\bi\beta}
      + \text{H.c.} \right)\\
    &-\sum_{\alpha,\beta}
    \left(i V_{-\nu} \sigma^{\gamma}_{\alpha\beta}
      b_{\bi+\benu\alpha}^{\dagger}a^{\phantom\dagger}_{\bi\beta}
      + \text{H.c.} \right).
  \end{split}
\end{equation}
The global current is defined as the sum over all bonds/sites
\begin{equation}
  J^{\gamma}_{\nu} \equiv \sum_{\bi}j^{\gamma}_{\bi\nu}.
\end{equation}

\section{Extension of a result by Brillouin\label{sec:app2}}

In this section, we show that a state that carries a finite current of locally
conserved density cannot be a ground state. We follow the proof in
Ref.~\onlinecite{Ohashi1996}. Let us assume that $|\Psi\rangle$ is a
ground state, which has a finite expectation value of global current
$\langle \Psi | J^{\gamma}_{x} |\Psi\rangle=J\neq0$, and construct a state
\begin{equation}
  \ket{\Phi} \equiv \exp(-i\delta X^{\gamma}) \ket{\Psi},
\end{equation}
where
\begin{equation}
  X^{\gamma} \equiv \sum_{k,l} k s^{\gamma}_{(k,l)}.
\end{equation}
Since $X^{\gamma}$ commutes with $H_{\text{loc}}$ we get
\begin{equation}
  \label{eq:expval}
  \begin{split}
    \matrixel{ \Phi}{ H}{\Phi} &=
    \matrixel{ \Psi }{ H_{\text{loc}} }{\Psi} + \\
    \sum_{\nu} &\matrixel{ \Psi}{
      \exp(i\delta X^{\gamma})(T_{\nu} + T_{\nu}^{\dagger}) \exp(-i\delta X^{\gamma})}{\Psi}.
  \end{split}
\end{equation}
The operators $X^{\gamma}$ and $T_{\nu} + T_{\nu}^{\dagger}$ are
Hermitian, we can thus expand (\ref{eq:expval}) using the
Baker-Hausdorff lemma
\begin{equation}
  \begin{split}
    \label{eq:BH_lemma}
    &\exp(i\delta B)A\exp(-i\delta B)
    =A + i \delta [B, A]+ \\
    &\frac{(i\delta)^2}{2!} [B,[B,A]] +
    \dots\frac{(i\delta)^n}{n!} [B,[B,\ldots,[B,A]\ldots]]\ldots
  \end{split}
\end{equation}

To compute $[X^{\gamma}, T_{\nu}]$ we use Eq.~\ref{eq:commut1} and obtain
\begin{align}
  [X^{\gamma}, &T_{y}]=0,\\ \nonumber
  [X^{\gamma}, &T_{x}]=-\sum_{k,l}k[T_{x},s^{\gamma}_{(k,l)}]\\ \nonumber
  =&-\sum_{\alpha\beta}
     \sigma^{\gamma}_{\alpha\beta}
     \sum_{k,l}\left(k-(k+1)\right) \\ \nonumber
  \times \biggl[ &t_a a_{(k+1,l)\alpha}^{\dagger}
                   a^{\phantom\dagger}_{(k,l)\beta} \biggr.
                   \biggl. +t_b b_{(k+1,l)\alpha}^{\dagger}
                   b^{\phantom\dagger}_{(k,l)\beta}  \biggr. \\ \nonumber
  \biggl. &V_x a_{(k+1,l)\alpha}^{\dagger}
            b^{\phantom\dagger}_{(k,l)\beta} +
            V_{-x} b_{(k+1,l)\alpha}^{\dagger}
            a^{\phantom\dagger}_{(k,l)\beta} \biggr] \\ \nonumber
  = \sum_{\bi} \sum_{\alpha\beta} \sigma^{\gamma}_{\alpha\beta}
  \biggl[ &t_a
            a_{\bi+\bex\alpha}^{\dagger}a^{\phantom\dagger}_{\bi\beta}
            + t_b b_{\bi+\bex\alpha}^{\dagger} b^{\phantom\dagger}_{\bi\beta}
            \biggr. \\ \nonumber
  \biggl. +&  V_x
             a_{\bi+\bex\alpha}^{\dagger} b^{\phantom\dagger}_{\bi\beta}+
             V_{-x} b_{\bi+\bex\alpha}^{\dagger}a^{\phantom\dagger}_{\bi\beta}
             \biggr] .\\ \nonumber
\end{align}
Using the identity $[X^{\gamma},T^{\dagger}]=-[X^{\gamma},T]^{\dagger}$ we arrive at
\begin{equation}
  \begin{split}
    &[X^{\gamma}, T_x + T_x^{\dagger}] 
    =iJ_x^{\gamma}.
  \end{split}
\end{equation}
We can also evaluate the next commutator
\begin{align} \nonumber
  &[X^{\gamma}, [X^{\gamma}, T_x + T_x^{\dagger}]] =
    [X^{\gamma}, i J_x^{\gamma}] \\ \nonumber
  \nonumber
  =\sum_{k,l} &\qty( k+1-k) \\
  \nonumber
  \times \sum_{\alpha} &\left[ t_a
                         a_{(k+1,l)\alpha}^{\dagger}a^{\phantom\dagger}_{(k,l)\alpha}
                         +t_b
                         b_{(k+1,l)\alpha}^{\dagger}b^{\phantom\dagger}_{(k,l)\alpha}
                         \right.\\ \nonumber
  &\left. + V_x
    a_{(k+1,l)\alpha}^{\dagger}b^{\phantom\dagger}_{(k,l)\alpha}
    + V_{-x}
    b_{(k+1,l)\alpha}^{\dagger}a^{\phantom\dagger}_{(k,l)\alpha}
    \right]\\ \nonumber
  &+ H.c. \\ 
  =T_x&+T_x^{\dagger}.
\end{align}

We finally obtain
\begin{equation}
  \begin{split}
    \matrixel{\Phi}{H}{\Phi} &=
    \matrixel{\Psi}{H}{\Psi} - \sin\delta
    \matrixel{\Psi}{J_x^{\gamma}}{\Psi} \\
    &+(\cos \delta - 1)  \matrixel{\Psi}{T_x + T_x^{\dagger}}{\Psi}\\
    &=\langle\Psi|H|\Psi\rangle - \delta J + \mathcal{O}(\delta^2).
  \end{split}
\end{equation}
Therefore $\ket{\Psi}$ cannot be a ground state if $J$ is
finite.

\printendnotes

\end{document}